\documentstyle[preprint,aps]{revtex}

\begin{document}
\draft

\tighten
\preprint{\vbox{
\hbox{TUTP-95-5}
\hbox{SISSA Ref.158/95/A}
\hbox{CAT-95/05}
hep-ph/9512290}}

\title{BUBBLE COLLISIONS AND DEFECT FORMATION \newline
IN A DAMPING ENVIRONMENT}
\author{Antonio Ferrera \thanks{%
e-mail: aferrera@diamond.tufts.edu}}
\address{Physics Department, Tufts University. Medford, MA 02155.}
\author{Alejandra Melfo \thanks{%
e-mail: melfo@stardust.sissa.it}}
\address{International School for Advanced Studies, 34014 Trieste, Italy,
{\rm and}\\ Centro de Astrof\'{\i}sica Te\'orica, Universidad de Los
Andes, M\'erida 5101-A, Venezuela}
\date{10/12/95}\tighten

\maketitle

\begin{abstract}
Within the context of a first-order phase transition in the early Universe,
we study the collision process for vacuum bubbles expanding in a plasma. The
effects of the plasma are simulated by introducing a damping term in the
equations of motion for a $U(1)$ global field. We find that
Lorentz-contracted spherically symmetric domain walls adequately describe
the overdamped motion of the bubbles in the thin wall approximation, and
study the process of collision and phase equilibration both numerically and
analytically. With an analytical model for the phase propagation in 1+1
dimensions, we prove that the phase waves generated in the bubble merging
are reflected by the walls of the true vacuum cavity, giving rise to a
long-lived oscillating state that delays the phase equilibration. The
existence of such a state in the 3+1 dimensional model is then confirmed by
numerical simulations, and the consequences for the formation of vortices in
three-bubble collisions are considered.
\end{abstract}


\section{Introduction}

According to the standard model and its extensions, symmetry breaking phase
transitions are expected to occur in the early universe. The mechanism by
which these transitions may take place could be either by formation of
bubbles of the new phase within the old one (i.e., first order phase
transition) or by spinodal decomposition (i.e.,second order phase
transition). Although the verdict is still pretty much in the air as to
which one would have actually taken place, in the particular case of the
electroweak phase transition common opinion inclines more towards the first
of the two possibilities. As this scenario would have it, bubbles of the new
phase nucleated within the old one (the nucleation process being described
by instanton methods as far as the WKB approximation remains valid \cite{c77}%
), and subsequently expanded and collided with each other until occupying
basically all of the space available at the completion of the transition.
The caveat introduced by the word ``basically'' is important though: in the
process of bubble collision, the possibility arises that regions of the old
phase become trapped within the new one, giving birth to topologically
stable localized energy concentrations known as {\it topological defects }
(for recent reviews see refs. \cite{td-rev} ), much in the same way in which
these structures are known to appear in condensed matter phase transitions.
  From  a theoretical point of view, topological defects will appear whenever a
symmetry group $G$ is spontaneously broken to a smaller group $H$ such that
the resulting vacuum manifold $M=G/H$ has a non-trivial topology: cosmic
strings for instance, vortices in two space dimensions, will form whenever
the first homotopy group of $M$ is non-trivial, i.e., $\pi _1(M)\neq 1.$

To see how this could happen in detail, let's consider the Lagrangian

\begin{equation}
{\cal L}=\frac 12\partial _\mu \Phi \partial ^\mu \Phi ^{*}-V(\Phi )
\label{lag}
\end{equation}
for a complex field $\Phi $. If $V$ is a potential of the type $V=\frac
\lambda 4(|\Phi |^2-\eta ^2)^2$ and its parameters are functions of the
temperature such that at high temperatures $\Phi =0$ is the only minimum of $%
V$, while at zero temperature all the $|\Phi |=\eta $ values of the field
correspond to different minima, then the structure of the vacuum manifold
will be that of $S^1$. As $\pi _1(S^1)\neq 1$, cosmic strings should then be
created in this model during the phase transition.

The way in which this would actually take place is via the Kibble mechanism
\cite{k76}. The basic idea behind it is that when two regions in which the
phase of the field takes different values encounter each other, the phase
should interpolate between this two regions following a geodesic in the
vacuum manifold. In the context of a first order phase transition, a
possible scenario for vortex formation in two space dimensions would then
look like this: three bubbles with respective phases of $0,$ $2\pi /3,$ and $%
4\pi /3$ collide simultaneously. According to the so-called geodesic rule
then, if we walk from the first bubble to the second , then from the second
to the third, and finally back from the third to the first again, the phase
will have wounded up by $2\pi $ in our path, having traversed the whole of
the vacuum manifold once along the way. Continuity of the field everywhere
inside the region contained by our path demands then that the field be zero
at some point inside of it, namely the vortex core. In the limit in which
the bubbles extend to infinity, outwards from the center of collision,
removal of this vortex would cost us an infinite amount of energy, since it
would involve unwinding the field configuration over an infinite volume. The
vortex is thus said to be topologically stable. In three space dimensions,
the resulting object would obviously be a string, rather than a vortex.

Clearly, there are other ways in which strings could be formed. Collisions
of more than three bubbles could also lead to string formation, or, for
instance, two of the bubbles could hit each other first, with the third one
hitting only at some later time while the phase is still equilibrating
within the other two. This event in particular will be far more likely than
a simultaneous three way (or higher order) collision, and it is probably the
dominating process by which strings are formed (especially if nucleation
probabilities are as low as required for WKB methods to be valid). If this
is the case, the importance of understanding the processes that take place
when two bubbles collide, as well as the need to clarify the role of the
third bubble and the conditions that should be met in order to get vortex
formation, becomes clear.

Bubble collisions have been studied by a number of groups, most notably by
Hawking {\it et al.} \cite{hms82}, Hindmarsh {\it et al. } \cite{hdb94},
Srivastava \cite{s92} and more recently by Melfo and Perivolaropoulos \cite
{mp95}, but important aspects of the question remain unclear so far. For
instance only in \cite{s92} and \cite{mp95} the posterior collision of a
third bubble is treated (the first two references studied only two bubble
collisions), and in none of the referred works did the authors concern
themselves with the interaction between the bubble field and the surrounding
plasma. From current work in the electroweak phase transition however, we
can expect that in many cases such interaction will not be negligible (e.g,
see Ref \cite{ew}). The basic underlying reason is relatively simple to
understand: as the bubble wall sweeps through an specific point, the Higgs
field $\Phi $ acquires an expectation value, and the fields coupled to it
acquire a mass. Thus, particles with not enough energy to acquire the
corresponding mass inside the bubble will bounce off the wall (thus
imparting negative momentum to it), while the rest will get through.
Obviously, the faster the wall propagates the stronger this effect will be,
since the momentum transfer in each collision will be larger, and thus a
force proportional to the velocity with which the wall sweeps through the
plasma appears.

The aim of this paper will thus be to explore two bubble collisions in an
environment that damps the bubble motion to try and establish how phase
diffusion and equilibration will proceed in this case, both analytically and
numerically. The damping that we will consider here will only affect the
modulus of the field though (which is equivalent to damping only the bubble
wall motion), since, the phase being a Goldstone boson, we do not expect it
to couple to the plasma as strongly as the modulus. We will find significant
differences with the undamped case, the main new features being related to
the possibility of the phase wave catching up with the bubble wall and
bouncing off it. Once that is understood, we will study collisions with a
third bubble and see under what circumstances we can expect to have vortex
formation. All this will be done in the context of a global $U(1)$ symmetry
breaking model.

The paper is organized in three sections: in the first one we establish the
general analytical model, in the second we study two bubble collision and
phase bouncing in 1+1 and 2+1 dimensions (the 3+1 case should be analogous
to the latter one), and finally the third is devoted to three bubble
collisions in 2+1 dimensions.

\section{Analytical model}

Consider the Lagrangian (\ref{lag}) for a complex field $\Phi $. We will use
the same form of potential that was used in \cite{s92,mp95}, that is,
\begin{equation}
V=\lambda \left[ \frac{|\Phi |^2}2\left( |\Phi |-\eta \right) ^2-\frac
\epsilon 3\eta |\Phi |^3\right] .  \label{pot}
\end{equation}
This is just a quartic potential with a minimum at $|\Phi |=0$ (the false
vacuum), and a set of minima connected by a $U(1)$ transformation (true
vacuum) at $|\Phi |=\rho _{tv}\equiv \frac \eta 4(3+\epsilon +\sqrt{\left(
3+\epsilon \right) ^2-8})$, towards which the false vacuum will decay via
bubble nucleation. It is the dimensionless parameter $\epsilon $ that is
responsible for breaking the degeneracy between the true and the false vacua.

The equations of motion for this system are then
\begin{equation}
\partial _\mu \partial ^\mu \Phi =-\partial V/\partial \Phi .  \label{em}
\end{equation}
For the potential (\ref{pot}), approximate solutions of (\ref{em}) exist for
small values of $\epsilon$, the so-called thin wall regime \cite{l83}, and
are of the form

\begin{equation}
|\Phi |=\frac{\rho _{tv}}2\left[ 1-\tanh \left( \frac{\sqrt{\lambda }\eta }2%
(\chi -R_0)\right) \right] ,  \label{thin}
\end{equation}
where $R_0$ is the bubble radius at nucleation time and $\chi ^2=|\stackrel{%
\rightarrow }{x}|^2-t^2$. The bubble then grows with increasingly fast speed
and its walls quickly reach velocities of order 1. We are interested however
in investigating a model with damped motion of the walls due to the
interaction with the surrounding plasma. In order to model this effect, we
will insert a frictional term for the modulus of the field in the equation
of motion, namely
\begin{equation}
\partial _\mu \partial ^\mu \Phi +\gamma \stackrel{\cdot }{|\Phi |}%
e^{i\theta }=-\partial V/\partial \Phi ,  \label{damp-em}
\end{equation}
where $\stackrel{\cdot }{|\Phi |}\equiv $ $\partial |\Phi |/\partial
t,\theta $ is the phase of the field, and $\gamma $ stands for the friction
coefficient (which will as a matter of fact serve as parameter under which
we will hide our lack of knowledge about the detailed interaction between
the wall and the plasma). We have found that there are approximate
analytical solutions to (\ref{damp-em}) corresponding to one bubble
configurations that have reached their terminal velocity in the thin wall
limit . Solutions will be found in a constructive way: first we will suppose
that solutions for which the wall has the form of a traveling wave do exist
to find out what their terminal velocity would be, and then use this
expression for the velocity to obtain the detailed analytical form of such
solutions in the thin wall limit. Comparison with numerical simulations will
confirm the existence of such solutions. Writing $\Phi $ in polar form:
\begin{equation}
\Phi =\rho e^{i\theta },  \label{polar}
\end{equation}
the equation of motion (\ref{damp-em}) takes the form

\begin{mathletters}
\begin{eqnarray}
\partial _\mu \partial ^\mu \rho +\gamma \stackrel{\cdot }{\rho }
&=&-\partial V(\rho )/\partial \rho +2\rho \partial _\mu \theta \partial
^\mu \theta ,  \label{polar-em1} \\
\partial _\mu \left\{ \rho \partial ^\mu \theta \right\} &=&0.
\label{polar-em2}
\end{eqnarray}

For a single bubble configuration we take $\theta $ to be a constant (the
phase of the bubble). Equation (\ref{polar-em2}) is then automatically
satisfied and the equation left for the modulus of the field is

\end{mathletters}
\begin{equation}
\partial _\mu \partial ^\mu \rho +\gamma \stackrel{\cdot }{\rho }=-\partial
V(\rho )/\partial \rho .  \label{mod-em}
\end{equation}

To get an approximate expression for the terminal velocity of a bubble under
this equation of motion in the thin wall limit, note that because the wall
thickness is much smaller than the radius of the bubble, we can go to $1+1$
dimensions. Inserting then the ansatz $\rho =\rho (x-x_0(t))$ leads to

\begin{equation}
(1-\stackrel{.}{x}_0^2)\rho ^{\prime \prime }+(\stackrel{..}{x}_0+\gamma
\stackrel{.}{x}_0)\rho ^{\prime }=\partial V(\rho )/\partial \rho ,
\label{wall-em}
\end{equation}
where $\rho ^{\prime }\equiv \partial \rho /\partial x$. We then get an
effective equation for the wall motion simply by multiplying by $\rho
^{\prime }$ and integrating over $-\infty \leq x$ $\leq +\infty $ to get rid
of the extra degrees of freedom pertaining to the field, whence

\begin{equation}
(\stackrel{..}{x}_0+\gamma \stackrel{.}{x}_0)\left( \int_{-\infty }^{+\infty
}\rho ^{\prime 2}dx\right) =\int_{-\infty }^{+\infty }V^{\prime }dx=\Delta V,
\label{wall-em2}
\end{equation}
where $\Delta V$ is just the potential energy difference between the false
and the true vacuum phases, and we have used the fact that derivatives go to
zero far from the origin. The solution to (\ref{wall-em2}) for the initial
conditions $x_0(t=0)=R_{0,}\stackrel{.}{x}_0(t=0)=0$ is

\begin{equation}
x_0\left( t\right) =\frac 1\gamma \alpha t+\frac \alpha {\gamma ^2}
(e^{-\gamma t}-1)+R_0   \label{doce}
\end{equation}
where $\alpha \equiv \Delta V/\left( \int \rho ^{\prime 2}dx\right) $. Thus,
for values of $t\gg \gamma ^{-1}$ the bubble walls will have reached their
terminal velocity

\begin{equation}
v_{ter}=\frac{\Delta V}{\gamma \left( \int \rho ^{\prime 2}dx\right) }.
\label{term-vel}
\end{equation}
Although obtained in a slightly different way, this expression coincides
with what Heckler found in \cite{heckler}. Typical estimates for the
terminal velocities of thin wall bubbles in the electroweak phase transition
give $v_{ter}\sim 0.1$ \cite{ew} (although in recent work higher values have
been found, see \cite{mt95}).

To get an approximate expression for $\rho $ valid within this regime , it
suffices to rewrite (\ref{mod-em}) with a $\rho =\rho (r-r_0(t))$ ansatz,
where $r$ is the usual radial coordinate. Using $\stackrel{..}{r}_0=0$, $%
\stackrel{.}{r}_0\approx v_{ter}$, we get

\begin{equation}
(1-v_{ter}^2)\frac{\partial ^2\rho }{\partial r^2}+\left( \frac 2r+\gamma
v_{ter}\right) \frac{\partial \rho }{\partial r}=\partial V(\rho )/\partial
\rho ,  \label{asym-em}
\end{equation}
$r$ being the radial coordinate in 3 spatial dimensions. According to (\ref
{term-vel}) however the terminal velocity goes roughly like

\begin{equation}
v_{ter}\approx \frac{\Delta V}{\gamma \left( \frac{\rho _{tv}^2}{\delta _m^2}%
\delta _m\right) }=\frac{\Delta V\delta _m}{\gamma \rho _{tv}^2},
\label{roughvt}
\end{equation}
where $\delta _m$ is the bubble wall thickness and $\rho _{tv}$ is the true
vacuum value of the field. That is, at the values of $r$ for which the first
derivative of the field is important ($r\sim R$ for a thin wall bubble), we
have $R\gamma v_{ter}\sim \delta _m\ll 1,$ and the second sumand in the
parenthesis is negligible when compared to the first. Furthermore, since the
radius of thin-walled bubbles is very large, we can also neglect the term $%
\left( 2/r\right) \partial \rho /\partial r$ --this is just the standard
thin wall approximation. Thus, we are left with

\begin{equation}
(1-v_{ter}^2)\frac{\partial ^2\rho }{\partial r^2}=\partial V(\rho
)/\partial \rho ,  \label{asym-em2}
\end{equation}
whose solutions for the potential written in (\ref{pot}) will be

\begin{equation}
\rho =\frac{\rho _{tv}}2\left[ 1-\tanh \left( \frac{\sqrt{\lambda }\eta }2%
\frac{(r-v_{ter}t-R_0)}{\sqrt{1-v_{ter}^2}}\right) \right] ,  \label{wall}
\end{equation}
which is simply a Lorentz-contracted, moving domain wall.

We have followed closely the analytical study with numerical simulations of
the processes. This was done in two steps (for details see Ref \cite{mp95}):
first we find a numerical solution of the Euclidean equations of motion for
the field, whose analytic continuation into Minkowski space will give us the
shape of the instanton solution immediately after the tunneling has taken
place. The initial bubbles thus found were then placed in a two dimensional
lattice, and evolved according to (\ref{damp-em}) with a modified leapfrog
method and reflecting boundary conditions. In the following sections, we
will be comparing the analytical results with these simulations.

\section{Two bubble collisions: phase propagation and bouncing}

We will now try to take a look at the events that take place when two damped
bubbles collide. To do so, we will organize the section into three parts.
The first one will deal with the configuration and evolution of the phase
from the time when the bubbles are still very far away from each other to
the point when they have completed the collision process. The second one
takes on from there to study the propagation inside the bubbles of the phase
waves that originate at that time, and stretches out until the moment when
these waves catch up with the walls at the other end of the system. Finally,
the third part studies the interaction between the phase waves and the
bubble walls, finding how the former will bounce off the latter and
propagate again into the interior of the bubbles, which will thus behave
almost as a resonant cavity.

\subsection{Initial phase configuration}

Within the WKB approximation the bubble nucleation rate is small, and
therefore typical bubbles are nucleated a long distance apart from each
other. The approximate solution for a system of two bubbles is then simply
the sum of the two independent bubble solutions,

\begin{eqnarray}
\Phi (bubble1 &+bubble2)\equiv &\rho e^{i\theta } \simeq \Phi (bubble1)+\Phi
(bubble2)\equiv \rho _1e^{i\theta _1}+\rho _2e^{i\theta _2},  \label{apsol}
\end{eqnarray}
to exponential accuracy. From this it is easy to compute how the phase
interpolates between the two bubbles

\begin{equation}
\tan \theta =\frac{\rho _1\tan \theta _1+\rho _2\tan \theta _2}{\rho _1+\rho
_2},  \label{phasewall}
\end{equation}
which has a kink like shape centered at the origin. One could worry however
that, as the bubbles move towards each other, phase waves could be generated
in the central region that would subsequently propagate over the
exponentially small (but finite) background of the modulus so that (\ref
{phasewall}) would no longer adequately represent the combined phase of the
two bubbles in motion, even if (\ref{apsol}) still provided a good ansatz
for the combined modulus. To check whether this is indeed the case we can
solve the equation of motion for the phase (7b) using the modulus given by (%
\ref{apsol}). In a $1+1$ dimensional approximation, near $|x| = 0$ and far
away from the centers of each bubble we can take their moduli to be (we will
write simply $v$ for $v_{ter}$ now)
\begin{eqnarray}
\rho _1 &\simeq &\rho _{tv}e^{(x+vt-x_0)/\delta _m},  \label{veinte} \\
\rho _2 &\simeq &\rho _{tv}e^{-(x-vt+x_0)/\delta _m},  \nonumber
\end{eqnarray}
where we have taken the centers of the two bubbles to be initially situated
at $\pm x_0$. Then (\ref{apsol}) yields for the combined modulus
\begin{equation}
\rho ^2\simeq \rho _{tv}^2e^{2(vt-x_0)/\delta _m}\left\{ 2\cosh (2x/\delta
_m)+2\cos (\theta _1-\theta _2)\right\} .  \label{vuno}
\end{equation}
We can now use this in the equation of motion for the phase (\ref{polar-em2}%
), to get
\begin{equation}
\stackrel{..}{\theta }-\ \theta ^{\prime \prime }+\frac{2v}{\delta _m}%
\stackrel{.}{\theta }-\ \frac{2\delta _m^{-1}\sinh (2x/\delta _m)}{\cosh
(2x/\delta _m)+\cos (\theta _1-\theta _2)}\theta ^{\prime }=0.  \label{vtres}
\end{equation}
Note that for an initial phase difference $\theta _1-\theta _2=\pi $ the
modulus of the field (\ref{vuno}) is zero at the midpoint between the
bubbles. In this case then the denominator in the last term of (\ref{vtres})
goes to zero, which means that the phase has the shape of a step function as
we go over the origin, switching discontinuously from $\theta _1$ to $\theta
_2 $. On the other extreme, if the initial phase difference is zero there is
of course no dynamics to it. We will then focus in an intermediate
situation, and find solutions to (\ref{vtres}) for a phase difference of $%
\pi /2$. Using the ansatz $\theta =T(t)X(x)$, we can separate variables to
obtain:
\begin{mathletters}
\label{vcuatro}
\begin{eqnarray}
\stackrel{..}{T}+\ \frac{2v}{\delta _m}\stackrel{.}{T}+\ k^2T &=&0,
\label{vcuatroa} \\
X^{\prime \prime }+\frac 2{\delta _m}\tanh (\frac{2x}{\delta _m})X^{\prime
}+k^2X &=&0,  \label{vcuatrob}
\end{eqnarray}
$k$ being the separation constant. The equation for $T$ is obviously that of
a damped oscillator. For $k=0$, a $T=const.$ solution exists consistent with
our boundary condition that $\theta $ goes to $\theta _{1,2}$ as $%
x\rightarrow \pm \infty $ at all times, while, for $k\neq 0,$ solutions with
temporal dependence on the phase will die away on time scales of the order
of $\delta _m/v$ (as a matter of fact these solutions would seem to have
been artificially introduced by the switching on of the interaction between
the bubbles). To exponential accuracy then, the phase will rapidly settle in
a stationary state that will interpolate between its two values at infinity.
The solution to (\ref{vcuatrob}) for $k=0$ interpolating from say $\theta
_1=0$ at $x\rightarrow -\infty $ to $\theta _2=\pi /2$ at $x\rightarrow
+\infty $ will thus give us its complete behavior:
\end{mathletters}
\begin{equation}
\theta =\frac 12\arcsin (\tanh (\frac{2x}{\delta _m}))+\frac \pi 4.
\label{vcinco}
\end{equation}
Thus, the near false vacuum state around the bubbles behaves as a rather
effective insulating material as far as conduction of phase waves is
concerned, and we should not expect any significant propagation of the phase
until the bubbles run into one another. It may also be worth noting how the
phase interpolates from one of its values to the other by means of a
wall-like structure similar to walls created by the modulus of the field.
The thickness of this phase wall, $\delta _{p\text{,}}$ seems to be closely
related to the thickness of the bubble wall $\delta _m$, although this
relation will obviously change as the phase difference between the bubbles
change.

Once the bubbles collide and their walls begin to merge, the phase of each
bubble will start propagating into the other one at the speed of light.
While this merging is taking place, its net effect will be to widen the
phase wall, so that its final thickness will be {\it at least} twice the
thickness of the bubble walls for relativistic bubbles.

Figure 1 shows the results of simulations of this process. We have plotted
the phase and the bubble walls for a 1-dimensional, 2 bubble configuration
with an initial phase difference of $\pi /2$, for three different times of
the evolution: when they are about to start colliding, at the middle of the
merging process, and at the point of its completion respectively. We can
clearly see how the phase wall smoothens out and thickens while this takes
place.

\subsection{Phase propagation inside the bubbles}

Once the bubble walls have completed their merging, the phase is free to
propagate in the resulting single cavity. Since, especially in the thin wall
case, the modulus of the field inside the cavity remains essentially
constant and equal to its true vacuum value, the equation of motion for the
phase simplifies to a wave equation

\begin{equation}
\partial _\mu \partial ^\mu \theta =0.  \label{vsiete}
\end{equation}

Kibble and Vilenkin \cite{kv95} have studied the process of wave propagation
for the Abelian gauge model without dissipation. In the case they studied
the bubbles move essentially at the speed of light, and the problem has a
high degree of symmetry: if the axes are chosen so that the bubbles nucleate
along the $z$ axis at say $(0,0,0,\pm R)$, the whole bubble collision
process will be invariant under the 3-dimensional Lorentz group SO(1,2) in
the $(t,x,y)$ subspace. The bubble collision occurs then along the surface $%
z=0,\ t^2-x^2-y^2=R^2$, and for any point in that surface there will be a
frame of reference in which that is the point of first contact. Thus,
symmetry dictates that$\ \theta $ be a function
\begin{equation}
\ \ \theta =\theta (\tau ,z),  \label{vocho}
\end{equation}
where $\tau ^2=\ t^2-x^2-y^2$. In our case, the damped motion of the walls
obviously breaks Lorentz invariance as far as the bubbles motion and
collision is concerned. The points at which the two bubbles connect to each
other in the $z=0$ plane however will still move at a very high speed (in
fact much higher than the terminal velocity of the walls), and, since the
phase is not affected by the damping, SO(1,2) Lorentz invariance should
still be a good approximation in our case. Inserting the ansatz (\ref{vocho}
) for $\theta $ in (\ref{vsiete}) yields then the equation for the phase
\begin{equation}
\partial _\tau ^2\theta +\frac 2\tau \partial _\tau \theta -\partial
_z\theta =0.  \label{vnueve}
\end{equation}
As initial setup we will assume that the bubbles have been nucleated at $\pm
R$ in the $z$ axis while the collision takes place at $t=0$. For the initial
conditions we will take the thickness of the phase wall to be negligible
when compared to the other scales of the problem, that is,
\begin{equation}
\theta |_{\tau =0}=\theta _0\varepsilon (z)\;,\;\;\;\;\partial _\tau \theta
|_{\tau =0}=0,  \label{treinta}
\end{equation}
where $\varepsilon $ is the step function. Equation (\ref{vnueve}) can then
be solved to get
\begin{equation}
\theta =\left\{
\begin{array}{c}
\theta _0\;\qquad \text{ for\qquad }z>\sqrt{t^2-x^2-y^2}, \\
\theta _0\frac z{\sqrt{t^2-x^2-y^2}}\text{\quad for \ \ }|\,z\,|\leq \sqrt{%
t^2-x^2-y^2}, \\
-\theta _0\;\qquad \text{ for\qquad }-z>\sqrt{t^2-x^2-y^2}.
\end{array}
\right.   \label{tuno}
\end{equation}
This approximation will obviously work better for bubbles that move with
relatively high terminal velocities. Firstly of course because the Lorentz
invariance approximation will be more adequate, but also because faster wall
motion leads to smaller phase wall thicknesses, as seen at the end of the
last subsection, so that (\ref{treinta}) is a better approximation for the
phase wall. It will also clearly work better in the central region of the
bubble than close to its walls, where the interaction between these and the
phase wave, as well as the detailed way in which the connecting
circumference between the two bubbles expands, may be of importance.

To see how well this approximation holds we can compare it with numerical
simulations, which will however be performed in $2+1$ dimensions. The
solution to the $2+1$ version of (\ref{vnueve}) is easily found

\begin{equation}
\theta =\left\{
\begin{array}{c}
\theta _0\;\qquad \text{ for\qquad }z>\sqrt{t^2-x^2}, \\
\frac{2\theta _0}\pi \arcsin (\frac z{\sqrt{t^2-x^2}})\text{\quad for \ \ }%
|\,z\,|\leq \sqrt{t^2-x^2}, \\
-\theta _0\;\qquad \text{ for\qquad }-z>\sqrt{t^2-x^2},
\end{array}
\right.   \label{ttres}
\end{equation}
and has general features similar to those of the $3+1$ case. Figure 2a shows
a couple of snapshots of the phase wave profile along the $z$ axis at
different times after the collision for thin wall bubbles with relatively
low friction. As we can see, our results give a rather accurate picture of
how the phase evolves in this case. Figure 2b shows the same situation for
the case in which the bubbles move under high friction (i.e., low terminal
velocity). As we can see, here the phase wall thickness at the the time at
which the bubbles finish merging is rather large, and therefore the step
function approximation in (\ref{treinta} ) fails to correctly represent the
initial state of the phase. We should note however that the phase
propagation seems to proceed along similar lines as before, so that if we
were to solve (\ref{vnueve}) with a smoother ansatz for $\theta (\tau =0)$
we should again reproduce the observed phase behavior.

\subsection{Interaction between the phase wave and the bubble walls: phase
bouncing}

The most salient feature of the problem of bubble collision in the damped
regime will be the possibility of interaction between the phase waves that
propagate inside the merged bubbles, and the bubble walls. Such a situation
was of course never encountered in the case of walls moving in vacuum and
reaching asymptotically the speed of light.

We can roughly anticipate the outcome of this interaction on physical
grounds: since the phase is massless inside the bubbles, but massive in the
near false vacuum outside them, it should follow that only the contributing
modes to the phase wave which have sufficient energy to acquire the required
mass will be able to go through the wall at all, while all the others must
bounce off it. Whether these modes will exist at all in the phase wave, and
if they do in what proportion, will then determine how the phase wave will
behave after the bouncing.

We will first study the interaction between the phase wave and the bubble
wall both analytically and numerically in the simpler $1+1$ dimensional
scenario, and then try to extrapolate our conclusions to the $2+1$ case and
confront them with more numerical simulations there.

In $1+1$ dimensions, our damped thin wall solution looks like
\begin{equation}
\rho =\frac{\rho _{tv}}2\left[ 1-\tanh \left( \frac{\sqrt{\lambda }\eta }2%
\frac{\left( x-v_{ter}t-x_0\right) }{\sqrt{1-v_{ter}^2}}\right) \right] .
\label{tcuatro}
\end{equation}
Boosting to a frame of reference that moves along with the wall and has its
origin at its center leads then to
\begin{equation}
\rho =\frac{\rho _{tv}}2\left[ 1-\tanh \left( \frac{\sqrt{\lambda }\eta }2%
x\right) \right] .  \label{tcinco}
\end{equation}

We will be interested in the situation in which phase waves approach the
wall from $x\rightarrow -\infty ,t\rightarrow -\infty $ and look at the
asymptotic outcome for $t\rightarrow +\infty $, assuming that the incoming
phase waves carry a very small amount of energy as compared to that stored
in the wall so that we can neglect any back reaction on it due to the
collision. In these conditions we can then take the shape of the wall as
being essentially fixed by (\ref{tcinco}). The equation of motion for the
phase simplifies to
\begin{equation}
\stackrel{..}{\theta }-\theta ^{\prime \prime }-2\frac{\rho ^{\prime }}\rho
\theta ^{\prime }=0,  \label{tseis}
\end{equation}
or, doing a Fourier transform in $t$,
\begin{equation}
\theta ^{\prime \prime }+2\frac{\rho ^{\prime }}\rho \theta ^{\prime
}+\omega ^2\theta =0.  \label{tsiete}
\end{equation}
Using then (\ref{tcinco}) for $\rho $ as advertised and performing the
change of independent variable
\begin{equation}
y=\frac 1{1+e^{-\sqrt{\lambda }\eta x}}  \label{tocho}
\end{equation}
puts the phase equation in the form
\begin{equation}
y\left( 1-y\right) \theta ^{\prime \prime }+\left( 1-4y\right) \theta
^{\prime }+\frac{\omega ^2}{\lambda \eta ^2y\left( 1-y\right) }\theta =0,
\label{tnueve}
\end{equation}
where now the prime indicates derivatives with respect to $y$. Equation (\ref
{tnueve}) is a form of Papperitz equation with regular singular points at $%
0,1,+\infty ,$ (see Ref. \cite{kv95}) and consequently its solutions will be
given by hypergeometric functions.

To learn about the fate of an incoming wave we will look at the two
independent solutions of (\ref{tnueve}) around $y\rightarrow 1$ (i.e., $%
x\rightarrow +\infty $) and select the one that represents an outgoing wave.
In terms of $x$, these outgoing wave solutions are found to be ($B$ being an
arbitrary constant)
\begin{equation}
\theta _{x\rightarrow +\infty }\simeq \left\{
\begin{array}{c}
\left( -1\right) ^{\stackrel{\symbol{126}}{\omega }-1}B\,\;e^{x/\delta
_m}\;\;e^{ix\stackrel{\symbol{126}}{\omega }/\delta _m}\text{ ,}\qquad
|\omega |\delta _m>1 \\
\left( -1\right) ^{\stackrel{\symbol{126}}{\omega }-1}B\,\;e^{x/\delta
_m}\;\;e^{x\stackrel{\symbol{126}}{\omega }/\delta _m}\text{ ,}\qquad
|\omega |\delta _m<1
\end{array}
\right.  \label{cseis}
\end{equation}
with $\stackrel{\symbol{126}}{\omega }=\sqrt{1-\omega ^2/(\lambda \eta ^2)}$%
, and where for notational convenience we have used $\delta _m\equiv \left(
\sqrt{\lambda }\eta \right) ^{-1}$ for the bubble wall thickness. (Note that
the exponential growth of the solutions would apparently yield a diverging
energy density, specially for the case $|\omega |\delta _m<1$, more on this
later).

Using now the connection formulas for the hypergeometric functions, we can
express $\theta $ in terms of incoming and outgoing waves around $%
x\rightarrow -\infty $. We get

\begin{equation}
\theta _{x\rightarrow -\infty }\simeq \left( -1\right) ^{\stackrel{\symbol{%
126}}{\omega }-1}B\left[ \Gamma _1e^{i\omega x}+\Gamma _2e^{-i\omega
x}\right] .  \label{cnueve}
\end{equation}
where $\Gamma _1,\Gamma _2$ are combinations of gamma functions. Demanding
that the incoming wave has unit amplitude fixes the constant $B$ and
determines the reflection and transmission coefficients. We obtain as final
expressions for $\theta $
\begin{mathletters}
\label{cincuenta}
\begin{eqnarray}
\theta _{x\rightarrow -\infty } &\simeq &e^{i\omega x}+\frac{\Gamma _2}{%
\Gamma _1}e^{-i\omega x},  \label{cincuentaa} \\
\theta _{x\rightarrow +\infty } &\simeq &\left\{
\begin{array}{c}
\left( \Gamma _1\right) ^{-1}\;e^{x/\delta _m}\;\; e^{ix\stackrel{\symbol{126%
}}{\omega }/\delta _m}\text{ ,}\, |\omega |\delta _m>1 \\
\left( \Gamma _1\right) ^{-1}\,\;e^{x/\delta _m}\;\; e^{x\stackrel{\symbol{%
126}}{\omega }/\delta _m}\text{ ,}\, |\omega |\delta _m<1
\end{array}
\right. .  \label{cincuentab}
\end{eqnarray}
It only rests now to write down the precise form of the gammas and analyze
their behavior. We get

\end{mathletters}
\begin{equation}
\widetext \frac{\Gamma _2}{\Gamma _1}= \frac{\Gamma( 1+2i\omega \delta _m)
\Gamma( 2-\stackrel{\symbol{126}}{\omega }-i\omega \delta _m) \Gamma( -1-%
\stackrel{\symbol{126}}{\omega }-i\omega \delta _m) }{\Gamma( 1-2i\omega
\delta _m) \Gamma( 2-\stackrel{\symbol{126}}{\omega }+i\omega \delta _m)
\Gamma( -1-\stackrel{\symbol{126}}{\omega }+i\omega \delta _m) }
\label{ciuno}
\end{equation}
\narrowtext

Thus, for as long as $\stackrel{\symbol{126}}{\omega }$ remains real (i.e.,
for as long as $|\omega |\delta _m<1$), numerator and denominator in (\ref
{ciuno}) will be complex conjugate of each other, and the reflection
coefficient will be

\begin{equation}
R=\left| \frac{\Gamma _2}{\Gamma _1}\right| ^2=1.  \label{citres}
\end{equation}
When $|\omega |\delta _m>1,\,\stackrel{\symbol{126}}{\omega }\rightarrow i%
\stackrel{\symbol{126}}{\omega }$ however, and $R$ is
\begin{equation}
R=\frac{\sinh ^2\left( \pi \left( \omega \delta _m-\,\stackrel{\symbol{126}}{%
\omega }\right) \right) }{\sinh ^2\left( \pi \left( \omega \delta _m+\,%
\stackrel{\symbol{126}}{\omega }\right) \right) },  \label{cicuatro}
\end{equation}
and we see that $R\rightarrow 0$ for $\omega \delta _m\gg 1$. Thus incidents
waves with wavelength $1/\omega $ larger than the wall thickness will be
completely reflected by the wall, whereas waves with wavelengths shorter
than the wall thickness will be partially reflected and partially
transmitted, the reflected part going to zero quickly as the wavelength
decreases. Of course these conclusions will have to be somewhat modified to
describe the process from the point of view of a static observer at the
origin. For such a reference frame, a boost with velocity $-v_{ter}$ yields
for the new frequencies $\omega ^{\prime }$ of the incident and reflected
waves
\begin{eqnarray}
\omega _{in}^{\prime } &=&\beta \omega \left( 1+v_{ter}\right) ,
\label{cicinco} \\
\omega _{ref}^{\prime } &=&\beta \omega \left( 1-v_{ter}\right) =\omega
_{in}^{\prime }\frac{1-v_{ter}}{1+v_{ter}},  \nonumber
\end{eqnarray}
where $\beta \equiv 1/\sqrt{1-v_{ter}^2}.$ Thus, for a static observer the
condition $\omega \delta _m<1$ is transformed into
\begin{equation}
\omega _{in}^{\prime }\delta _m<\beta \left( 1+v_{ter}\right) ,
\label{ciseis}
\end{equation}
while the reflected wave will present a Doppler shift relative to the
incoming wave due to it having bounced off a moving wall.

Figures 3a, 3b show an incident wave train of wavelength roughly five times
that of the wall (in the static frame) being completely reflected. The
reflected wave has the same amplitude as the incident one, but its frequency
has decreased due to the Doppler shift. Figures 4a, 4b then show how a wave
train with wavelength about half the wall thickness is partially reflected.
Both cases present exponential growth of the phase outside the bubble
(remember that, modulo $2\pi $, the maximum phase difference that one can
have is $\pi $). This seems to fit the general behavior (\ref{cseis}). Note
however that an exponential growth of the phase with $x$ does not
neccesarily lead to an equivalent growth of the energy stored by the phase
gradient, since this goes like $\eta ^2\left( \partial \theta /\partial
x\right) ^2$ and the modulus is decaying exponentially. Thus, in the
expression for the case $\omega \delta _m>1$ the energy in the phase
gradient will tend to a constant for instance. If the behavior of the phase
outside the bubble were to be like the one given by the $\omega \delta _m<1$
regime in the same equation however, the phase gradient energy {\it would }%
diverge exponentially with $x$. Any of these two behaviours however will end
up in the breakdown of our approximation that the wall is unaffected by the
phase wave propagation. Once the phase wave gets through to the false
vacuum, the modulus of the field stops behaving like a transparent medium as
we saw in section III.1. In those conditions, and since the modulus decays
exponentially, no matter how small was the energy carried initially by the
phase wave we will always get to a point in which it will be of the same
order than the energy of the modulus. At that point, backreaction on the
modulus is no longer negligible, and a proper analysis of the problem would
require solving the two coupled equations.  The situation is depicted in
Figures 3c, 4c, where a blow up shows how the wall develops what appear as
small bumps in its decaying region. Figures 3d, 4d then, where the energy of
the modulus and of the phase are plotted, show how these bumps appear at the
points where the energy of the phase is getting larger than that of the
modulus. The result of this backreaction seems to be that the modulus
absorbs the excess energy in the phase, whereas the phase will still
propagate forward in some exponential way. We have not carried out a
detailed analysis of this part of the problem however, since it does not
directly affect the behavior of the phase inside the bubbles in any relevant
way.

To sum up, we can expect that waves with wavelengths (roughly) larger than
the bubble thickness will be totally reflected by the wall, the rest being
partially transmitted and partially reflected by it. It is then easy to see
what will happen to the phase wave in a ($1+1$ dimensional) two bubble
collision: as we saw in section III.1, the phase wall thickness at the end
of the collision will be {\it at least} of the order of twice the bubble
wall thickness, and more likely larger than that. It seems clear then that
basically all of the Fourier components of the phase wave will have
wavelengths that will fall into the total reflection regime, and thus the
whole phase wave itself will simply be reflected by the wall, while some
form of exponential behaviour propagates to the outside of the bubbles. Let
us imagine for the sake of clarity that, at collision time, bubble 1 had a
phase $\theta _1$ and bubble two $\theta _2>\theta _1$, with $\theta
_2-\theta _1=\Delta \theta $. If the shape of the phase wall at the
completion of the collision was $f(x)$ then, after it, we will have phase
waves with shape $f(x-t)/2,\,\,f(x+t)/2$ propagating into each bubble,
carrying a phase difference $-\Delta \theta /2$ into bubble 2 and $\Delta
\theta /2$ into bubble 1. After these waves have bounced off the bubble
walls and propagated back into the interior again, the phase of each bubble
will however be, for bubble 2, $\theta _2-2\Delta \theta /2=\theta _1$, and
for bubble 1, $\theta _1+2\Delta \theta /2=\theta _2$. The phases of the
bubbles will thus have switched. The whole process is depicted in Figure 5,
where the refered sequence has been plotted from a simulation. In Fig. 5a,
the walls of the two bubbles are just about to finish their merging
(continous line), and the shape of the phase wall at that time is shown
(dashed line). The bubble to the right plays the role of bubble 2 above,
having $\theta _2>\theta _1.$ The following pictures show how the two phase
waves propagate into the bubbles (Fig. 5b), and back after bouncing (Fig.
5c). As expected, after reflection each phase wave still carries a phase
of $\pm \Delta \theta /2,$ for bubbles 1 and 2 respectively. Finally, in
Fig. 5d the two phase waves meet again. The phase polarity of the system
has been completely inverted.

If this mechanism had no energy losses, the two bubbles would then behave as
a sort of resonator in which the phase of each bubble would continuously
oscillate between $\theta _1$ and $\theta _2,$their respective polarities
always switched. Since the bubble walls are moving though, the phase waves
present the expected Doppler shift after bouncing off the walls as was seen
before (which can also be clearly seen in Figs. 5c, d). For
non-relativistic terminal velocities (i.e., for $v_{ter}\ll 1$), the
magnitude of this shift is from (\ref{cicinco}) (we drop the primes in the
notation for the frequencies here)
\begin{equation}
\omega _{ref}\simeq \omega _{in}(1-2v_{ter}),  \label{cisiete}
\end{equation}
or what is the same, after $n$ reflections, the resulting phase wave will
have a frequency
\begin{equation}
\omega _{nref}\simeq \omega _{in}(1-2v_{ter})^n.  \label{ciocho}
\end{equation}
The oscillation process will effectively die off when $\omega _{nref}\sim (2%
{\cal R})^{-1}$, ${\cal R}$ being the (approximate) radius of the single
true vacuum cavity after $n$ reflections have taken place. If $\omega
_{in}^{-1}$ is very small when compared to ${\cal R}$, and neglecting higher
orders in $v_{ter}$, $\omega _{nref}$ becomes then of the order of the
radius after
\begin{equation}
n\simeq \frac 1{2v_{ter}}  \label{cinueve}
\end{equation}
oscillations. On the other hand, it is not difficult to see that the time
needed to complete each oscillation grows like
\begin{equation}
t_n=4R_0\left( \frac 1{1-v_{ter}}\right) ^n,  \label{sesenta}
\end{equation}
where $R_0$ is the radius of the bubbles at collision time (keeping in mind
that it will take only half $t_1$to reach the walls for the first time
though). Thus, the relaxation time until the oscillations die off will be
\begin{equation}
\tau _{osc}\simeq \frac{2R_0}{1-v_{ter}}+\stackrel{n\simeq 1/2v_{ter}}{%
\sum_{n=2}}4R_0\left( \frac 1{1-v_{ter}}\right) ^n.  \label{suno}
\end{equation}
For a terminal velocity around 0.1, (\ref{cinueve}) gives us five
oscillations until equilibration, for a total oscillating time $\tau
_{osc}\sim 24R_0$ from (\ref{suno}). We see therefore that the relaxation
time for the system can in general be quite significant.

\section{Three Bubble collisions: vortex formation}

We wish now to try to generalize the results found for phase wave bouncing
in 1+1 dimensions to 2+1 dimensions, since our final concern is to find
whether these effects can affect the formation of vortices. Finding an exact
solution for the phase propagation and interaction in 2+1 dimensions is an
extremely involved problem, but the results in one spatial dimensions can
certainly be used as a guide, and numerical simulations can provide the rest.

Note first that wave propagation in two or three spatial dimensions differs
qualitatively from the one dimensional case. Whereas in one dimension we
have two phase waves that propagate inside the bubbles carrying half of the
phase difference each, in two or more dimensions we have two wavefronts with
amplitudes that decrease in time and a region that continuously interpolates
between them (see the previous section). In Figure 6 we have plotted magnitude
and phase contours in a two-bubble collision, as well as the phase at each
point represented by an arrow. We can see how the phase interpolates between
the two bubbles at merging (6a), and then starts to propagate. During the
initial stages of propagation, the points of contact of the two bubbles move
in the direction perpendicular to the z axis at superluminal speed, so that
the phase waves cannot reach them. Propagation is similar in these first
stages to the undamped case: the region of interpolated phase is just a
semicircle (6b). Soon however the contact points will slow down to
asymptotically reach the terminal velocity. When the phase waves reach the
walls, the shape of the interpolating region is affected by the interaction,
as can be seen in 6c and d. We come then to the first consequence of the
damped motion of the walls. The fact that the phase waves are able to catch
up with the bubble walls means that, at this point, a region of smoothly
interpolated (i.e., nearly homogeneous) phase extends up to the boundaries
of the system. At this stage then it is easy for a third bubble of the right
size and position to collide with the merged two-bubble system {\em inside}
the region where the phase is nearly homogeneous. Hence, no vortex will be
formed. The situation is depicted in Fig. 7, where a third bubble collides
with the two bubble system while the processes seen in Figure 6 are taking
place. In 7a and b, the third bubble collides shortly after the phase starts
interpolating, so that a vortex is created. In 7c and d however the
nucleation of the third bubble occurs a little later. When it finally
reaches the two bubble system, the phase has already interpolated beyond the
collision points and no vortex is formed. Notice that the initial phases of
the bubbles in these figure are not exactly $\pi /6$, $5\pi /6$ and $3\pi /2$%
, but the two initial bubbles have slightly smaller values of the phases, so
that the resulting interpolated phase after merging is slightly smaller than
$\pi /2$. This was done in order to avoid having the third bubble colliding
with a region of {\em exactly opposite} phase. As was shown in Ref. \cite
{mp95}, such a collision produces a long-lived ``domain wall'' state between
the bubbles that delays the merging significantly. We will come back to this
point after we discuss the further evolution of the system below, for the
time being we just remark that the phase distribution is responsible for the
slight asymmetry in the vortex's position.

The situation depicted in Fig. 7 could represent a rather efficient
mechanism for the suppression of vortex formation, if it were not for the
interaction with the walls. From the 1+1 case, one expects the phase wave to
bounce back, thus spoiling its homogenization. How does the bouncing look
like in 2+1 dimension? In two spatial dimensions, at any point that the
phase wave meets the wall, its propagation vector will have one component in
the direction normal to the wall and another one tangential to it. Only the
component in the normal direction will see the wall and bounce off it
however, while the other one will continue to propagate freely. Thus,
although in general the interaction between the wall and the phase wave will
be a complicated superposition of these two processes, we can expect that
after some time, in the region close the walls the phase will predominantly
propagate tangentially to them, the rest of it having bounced towards the
center. The effects of this bouncing can already be seen in Figures 6c and
d: it is the reflection of the normal modes that causes the change of shape
in the wavefront.

After this we only have to wait for the central region of the phase,
propagating along the $z$ axis, to get to the end of the bubble and bounce
off the wall. Fig. 8 is a continuation of the same simulation started in Fig.
6. At the stage depicted by 8a, the phase wave has reached the end of the
bubble. Therefore, at that point virtually all the phase propagating normal
to the wall has bounced towards the center, and tangential propagation
dominates close to the wall. In Figure 8b we see the subsequent evolution of
the system. While the central region of the phase, propagating along the $z$
axis, collides head on and bounces off the end of the bubble as expected,
the tangential components cross each other at that region and start
propagating in the opposite direction, back towards the center of the
collision. Thus, the combination of these two phenomena brings about an
inversion of the phase similar to that found in 1+1 dimensions. Finally, a
third bubble collides with the system in 8c, {\em after} the bouncing has
taken place. Because of phase inversion, we get an antivortex formed in 8d.

Note that if we were to wait for the collision to happen until yet another
bounce had taken place (for a total of two bounces), the resulting defect
would again be a vortex. Three bounces and an antivortex again, and so on.
To get a total suppression of the vortex formation probability, therefore,
one would have to delay the collision of the third bubble until the system
has relaxed. As we have seen, the 1+1-dimensional model predicts a large
value for the relaxation time. Simulations show that for $v_{ter}\sim 0.1$,
vortices can still be formed in collisions occurring after the merged system
has gone through five oscillations, our predicted time for equilibration. So
the two-bubble system will have to remain in ``isolation'' for a very long
time to be able to acquire an homogeneous distribution of phase, a very
unlikely situation. Our results seem to indicate then that an analysis of
the vortex formation probability will have to take into account multiple
bubble collisions, such as for example the one carried out in Ref. \cite
{vvkb}

The situation is of course further complicated if we consider different
initial positions for the colliding bubbles. We have taken them to be
situated so that the latest bubble is centered in the plane of collision of
the other two, but it is easy to see that any shift in position will affect
the formation of vortices in a non trivial way, as it does any change in the
nucleation time. We conclude with some more words about the case depicted in
Fig. 7c and d. Had we set the three phases equally distributed, the delay
caused in the merging due to the formation of a metastable wall between the
region of opposite phases will have given enough time for the bouncing to
take place, and an antivortex to be formed. Simulations show that for the
positions of the bubbles considered here, a defect is {\em always} formed in
the particular (and less likely) case of equally distributed initial phases,
unless the phase oscillations have completely relaxed.

\section{Conclusions}

We have analyzed bubble motion and interactions in a plasma, simulating its
damping effect by a friction term in the equations of motion. We have found
that there exists an exact solution for the damped motion of the bubble in
the thin-wall regime, representing a bubble propagating with a terminal
speed roughly equal to the inverse of the friction coefficient.

An analytical study of the collision of two bubbles, and the subsequent
process of phase interpolation, was then performed. First, the shape of the
``phase wall'' that interpolates between the bubbles was found, and it was
shown that the vacuum state outside the bubbles is a very efficient
``insulator'', so that one does not expect the phase to propagate in the
false vacuum. The initial thickness of the phase wall was shown to be at
least twice that of the bubble's walls.

Once the initial set up for the collision was thus determined, the equations
for the phase propagation were solved, and the interaction with the walls
determined, using the previously found solution for the wall's motion. It
was shown that phase waves generated in a collision have wavelengths such
that they are always reflected by the walls. The result is that an
oscillating state is formed, in which the bubble's walls act almost as a
resonant cavity. Since the true vacuum cavity expands, though, these
oscillations are damped and eventually die off. For terminal velocities
typical of those expected in an electroweak phase transition, the relaxation
time is estimated to be around $24R$, with $R$ the radius of the bubbles at
collision.

Using numerical simulations, these results were shown to hold in 2+1
dimensions. The reflection of the phase wave goes along similar lines than
the one-dimensional case. The formation of vortices was shown to be affected
by the damping motion, basically due to two facts: first, that the
interpolated phase region can reach the boundaries of the system, where a
reflection of the phase waves off the bubble walls will take place; and
second that the long-lived oscillatory state thus produced delays
significantly the homogenization of the phase. The main effect of this
oscillatory state will have on vortex formation processes will be that it
will become possible, for the same set of three bubbles, to form a vortex,
and antivortex or no defect at all depending on the precise timing of the
last collision.

\section*{Acknowledgements}

This project was initiated by, and later enriched with, discussions with
Alex Vilenkin and Leandros Perivolaropoulos, to whom we wish to express our
gratitude. One of us (A.F.) would like to thank Alex Vilenkin for his
patience and countless suggestions.

\samepage

\begin{figure}[tbp]
\caption{Modulus of the field (a) and phase (b) of a system of two bubbles,
along the axis that joins their centers, for three moments of the collision
process. Continuos line is for $t=16 $, dotted for $t= 19 $, dashed for $t=
20 $ (where radial and time coordinates are given in units of the field's
mass, and the modulus is normalized with the symmetry-breaking scale).
Bubbles are nucleated at $t=0$, and $\gamma =3$. Here and in the following
graphs, $\epsilon= 0.8$. }
\label{fig1}
\end{figure}

\begin{figure}[tbp]
\caption{Phase propagation inside the bubbles. In (a), the friction
coefficient $\gamma$ is 3, in (b) is 10. The analytical solution is shown
for different times with a continuos line. Results of the simulation of the
phase propagation are shown with triangles (for $t= t_c + 1$, $t_c$=
collision time ) and circles (for $t= t_c + 7$). }
\label{fig2}
\end{figure}

\begin{figure}[tbp]
\caption{Interaction of a phase wave with the bubble wall, for a phase
wavelength approximately equal to five times the width of the wall. Arrows
indicate the direction of propagation of bubble and phase. In (a), (b) and
(c), the field's phase (dashed line) and magnitude (continuos line) are
plotted for different times as indicated. In (d), the energy in modulus
(continuos line) and phase (dashed line) are compared. }
\label{fig3}
\end{figure}

\begin{figure}[tbp]
\caption{Same as Fig. 3, where now the phase wave has a wavelength of
approximately half the wall's width.}
\label{fig4}
\end{figure}

\begin{figure}[tbp]
\caption{Phase propagation and bouncing. The phase (dashed line) and the
bubble walls (continous) are plotted for (a)$t=15$, (b)$t=25$, (c)$t=50$
and (d)$t=60$, with $\gamma = 10$.}
\label{fig5}
\end{figure}

\begin{figure}[tbp]
\caption{Collision of two bubbles with phases $\pi/6$ and $5 \pi/6$, and
subsequent phase interpolation. Continuos lines are contours of equal
modulus of the field, dotted lines are phase contours. The phase of the
field is also represented by arrows. In this and the following graphs, $%
\gamma = 10$. The axes are arbitrary lattice coordinates. }
\label{fig6}
\end{figure}

\begin{figure}[tbp]
\caption{Collision of three bubbles with phases $(\pi/6 - \Delta)$, $5 \pi/6
- \Delta)$ and $3 \pi/2$, with $\Delta = 0.01$. (a) and (b) are plots of two
moments of the collision of two bubbles nucleated at $t=0$ with a third one
nucleated at $t=8$. (c) and (d) correspond to a similar simulation, where
now the third bubble is nucleated at $t=18$, in the same position as the
previous one. Only field magnitude contours are represented. }
\label{fig7}
\end{figure}

\begin{figure}[tbp]
\caption{Later evolution of the system of Figure 6. A third bubble is
nucleated at $t=48$. Phase contours are omitted in the last frame for
clarity.}
\label{fig8}
\end{figure}

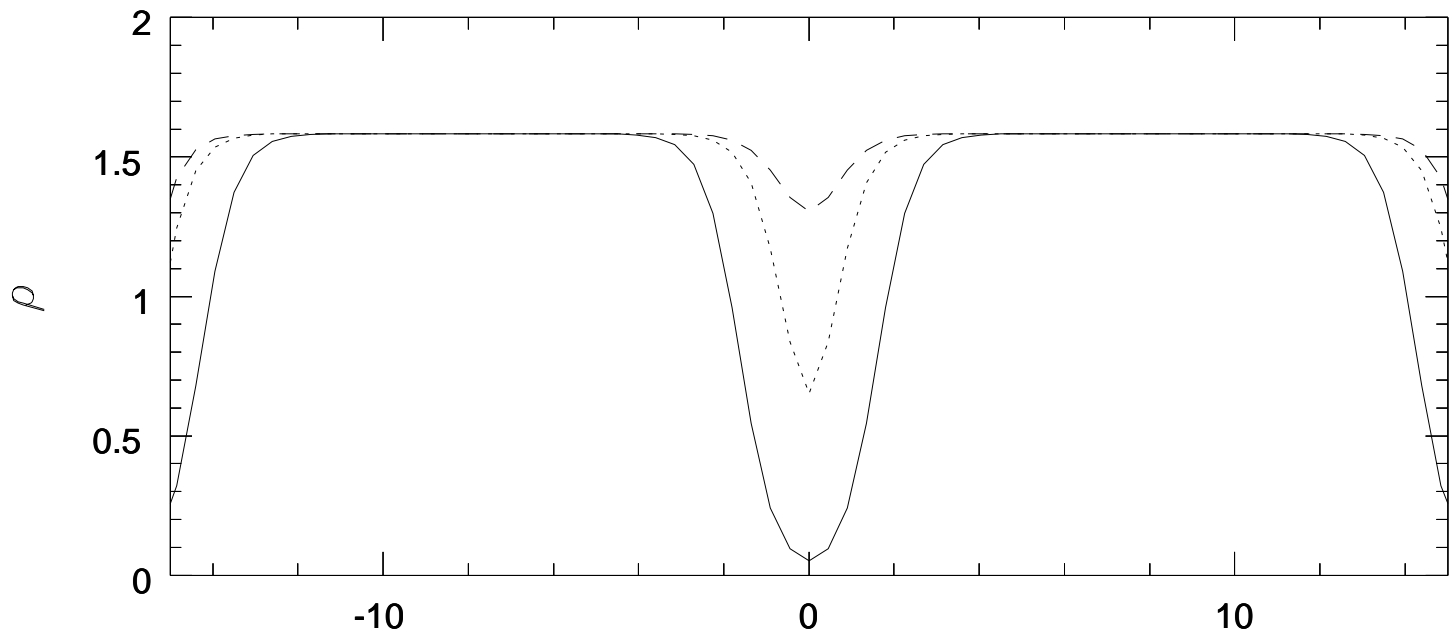
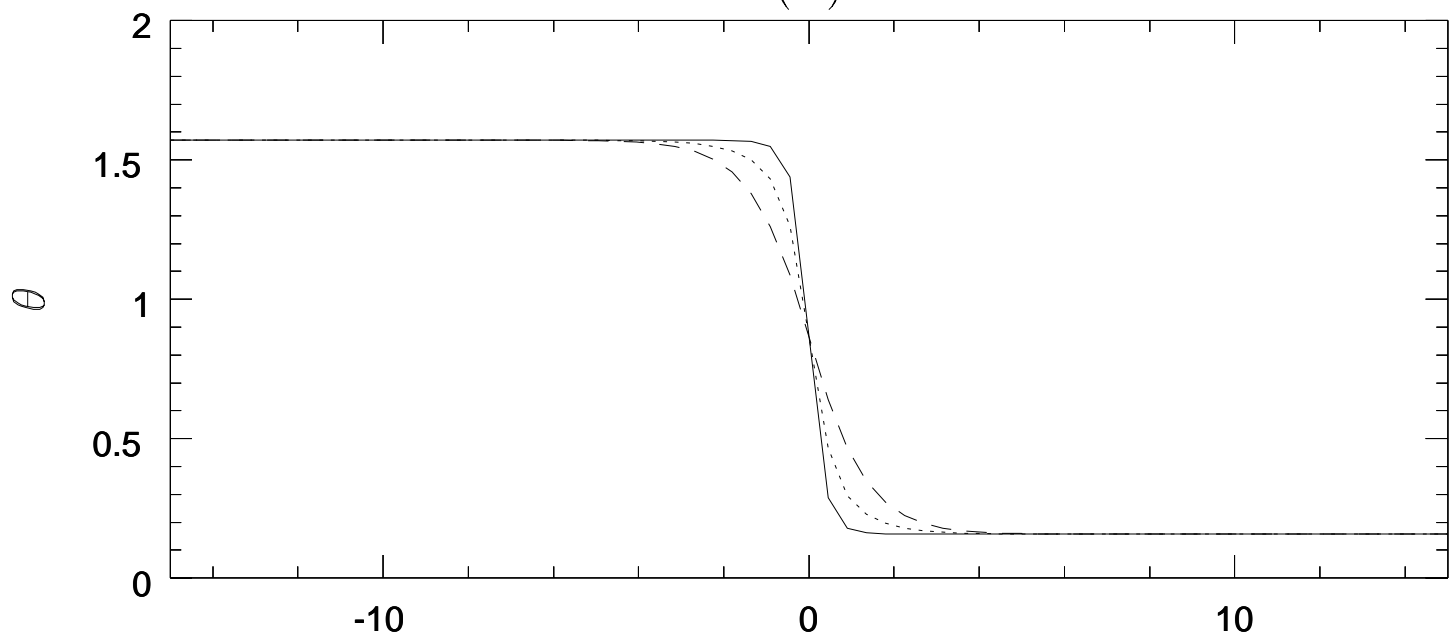
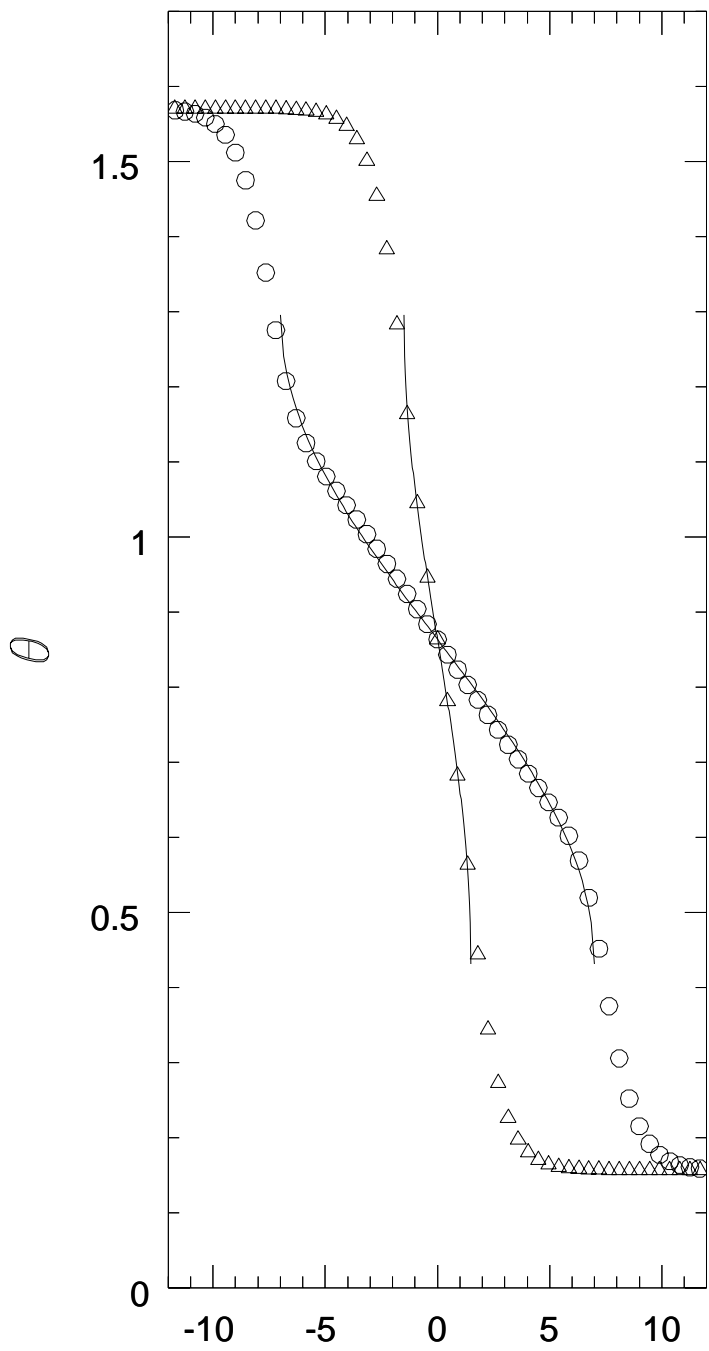
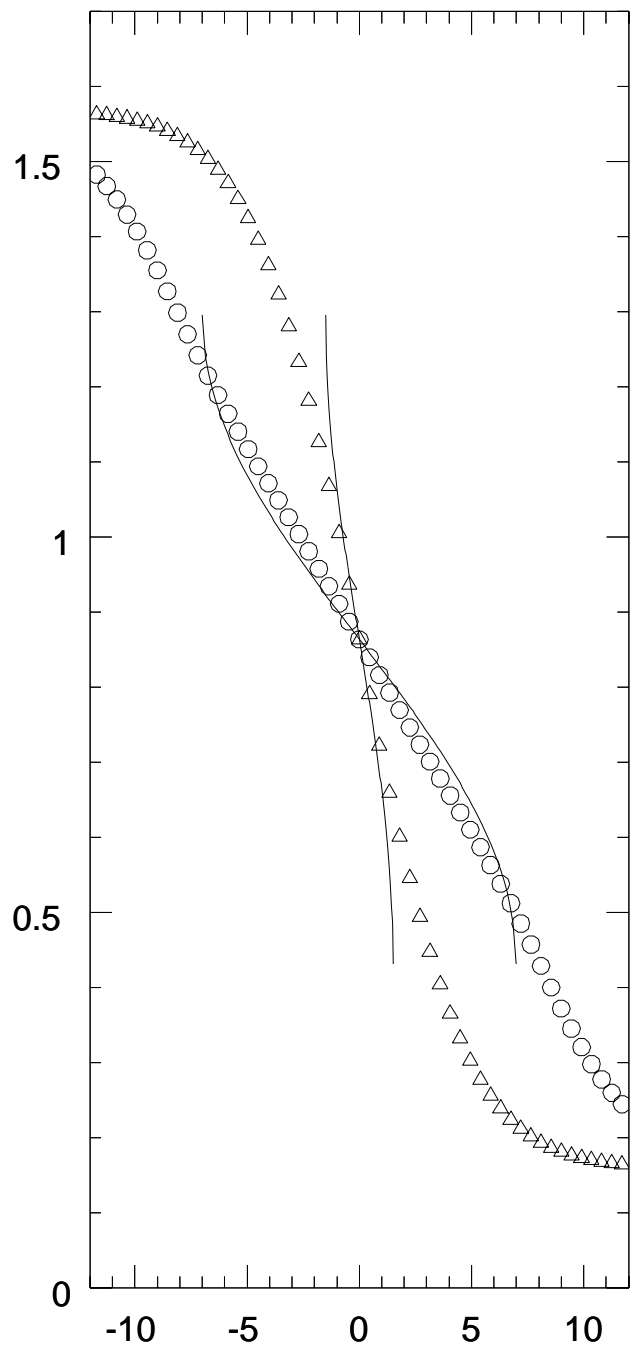
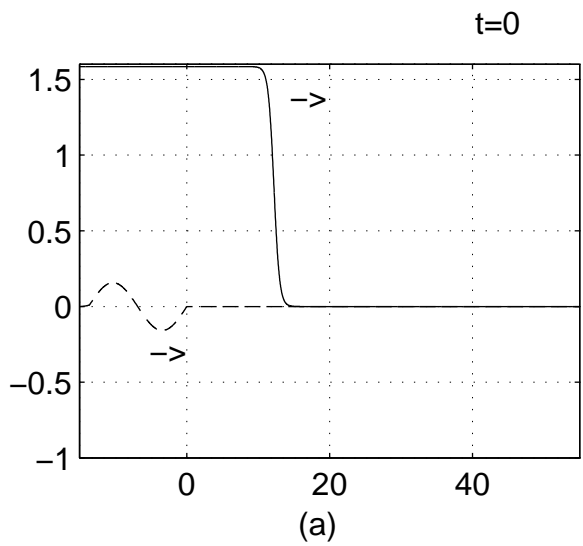
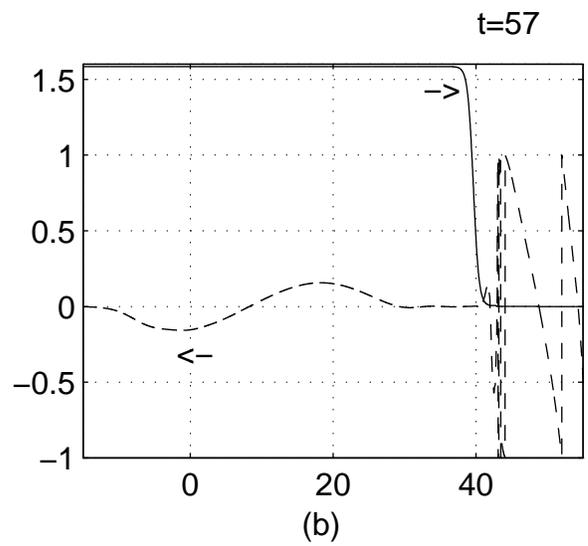
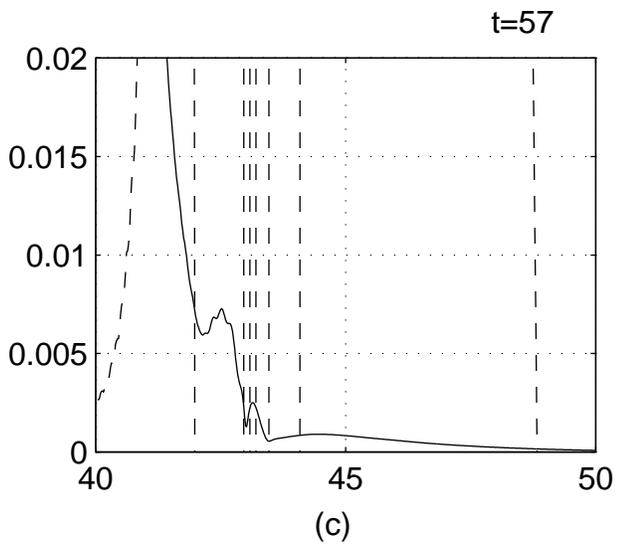
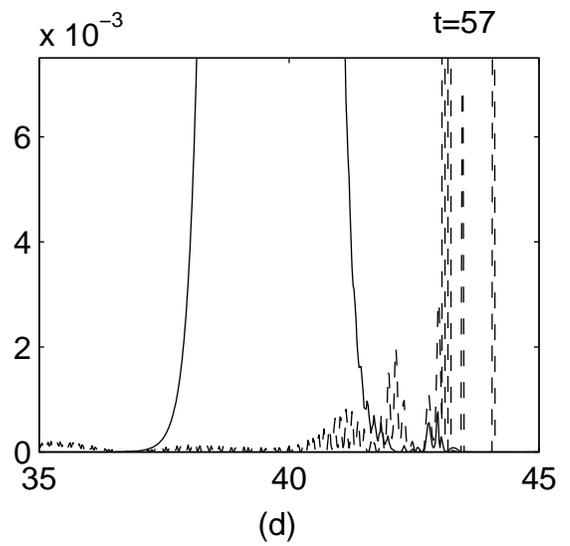
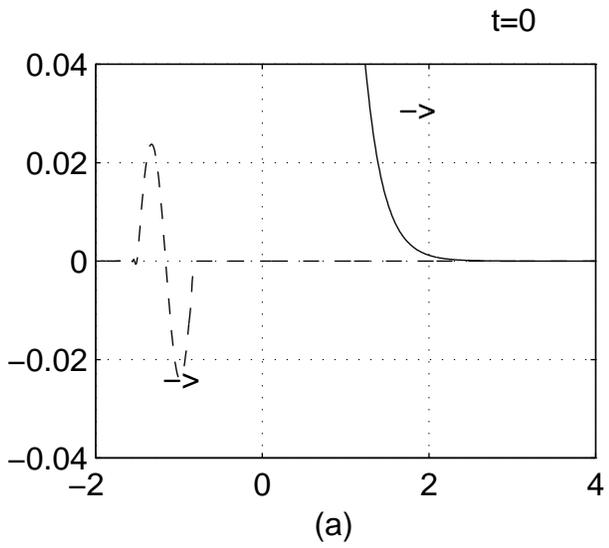
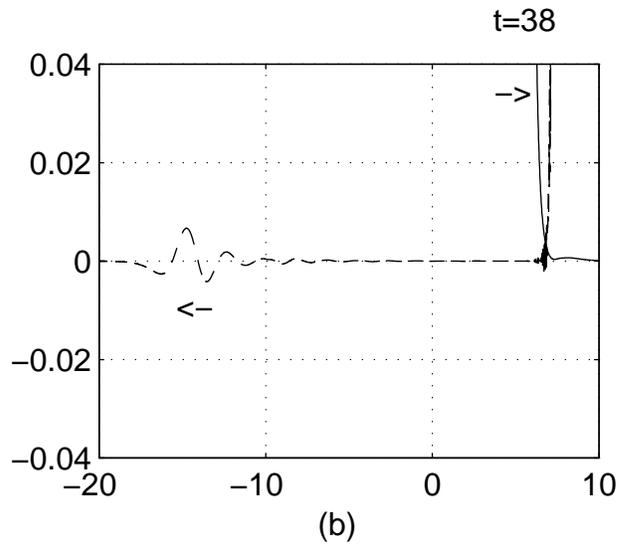
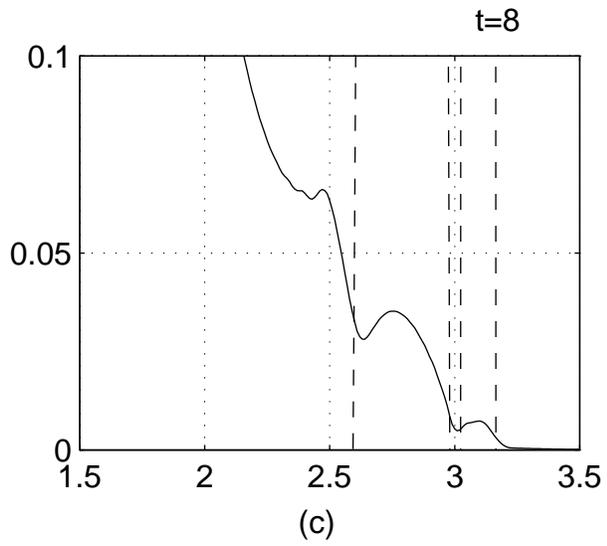
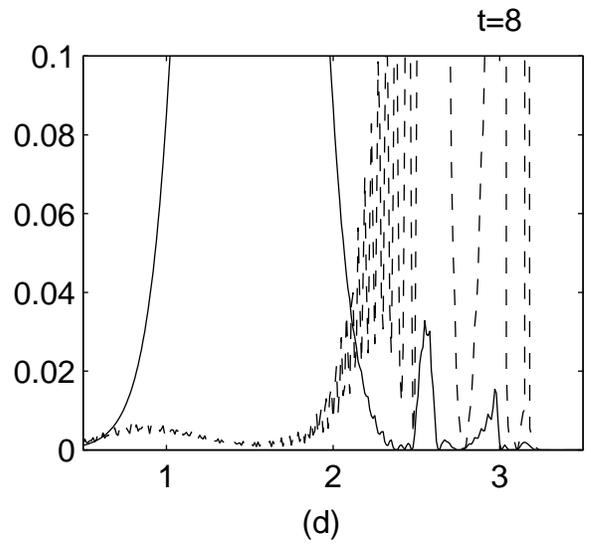
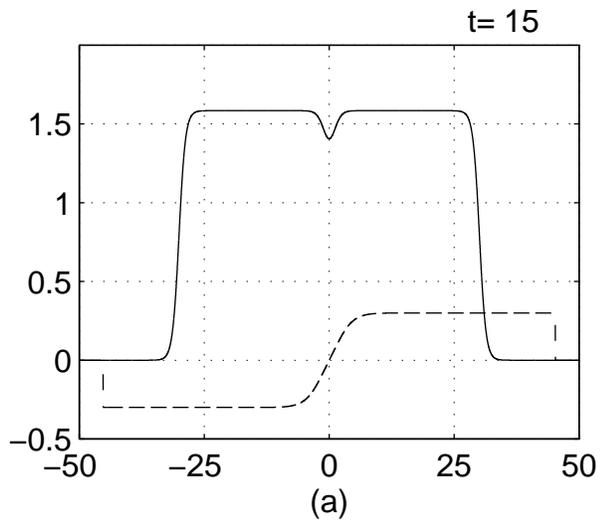
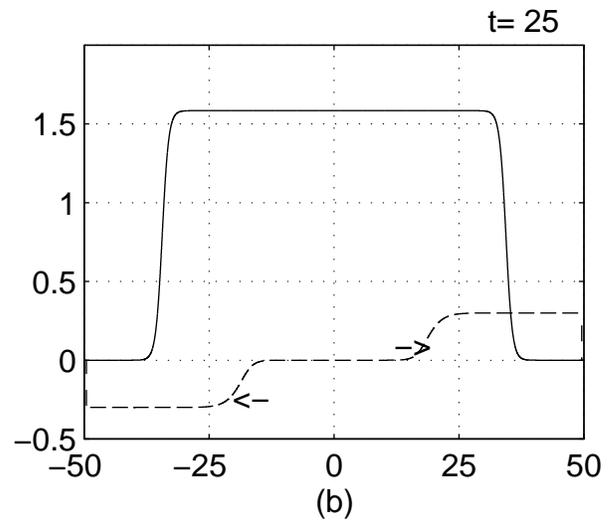
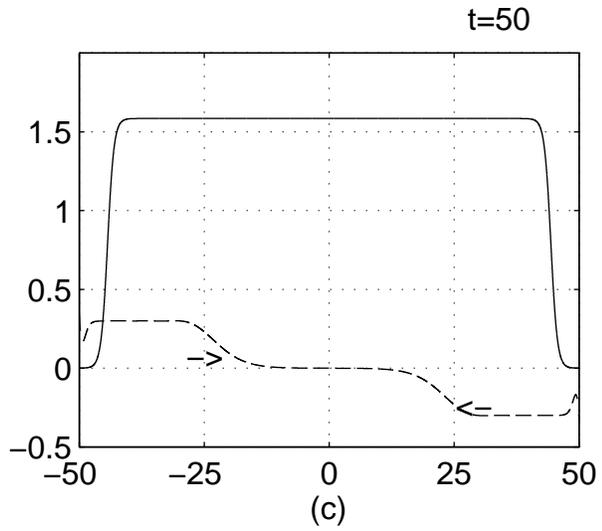
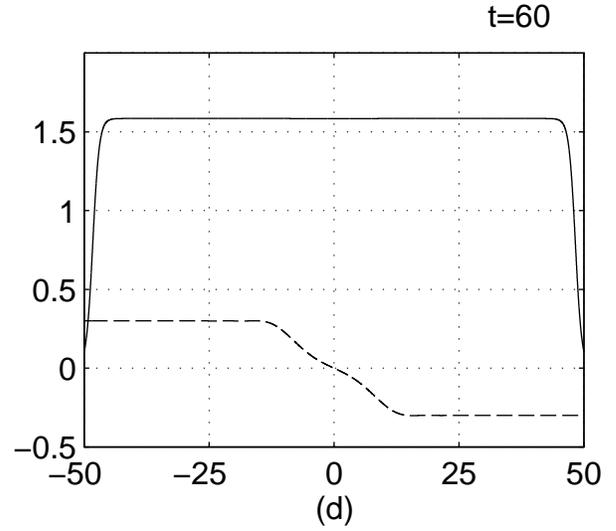
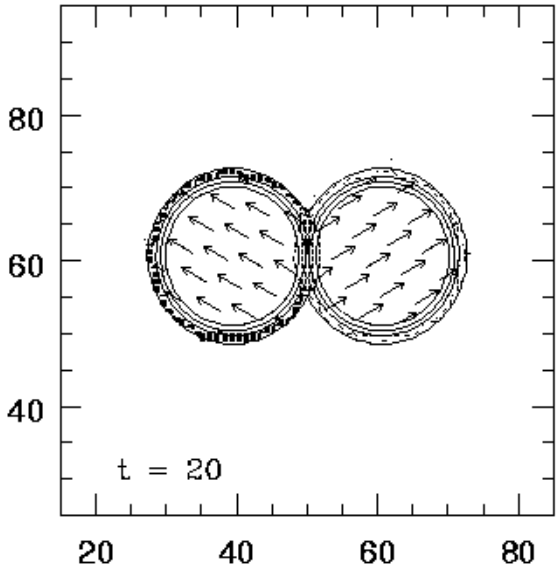
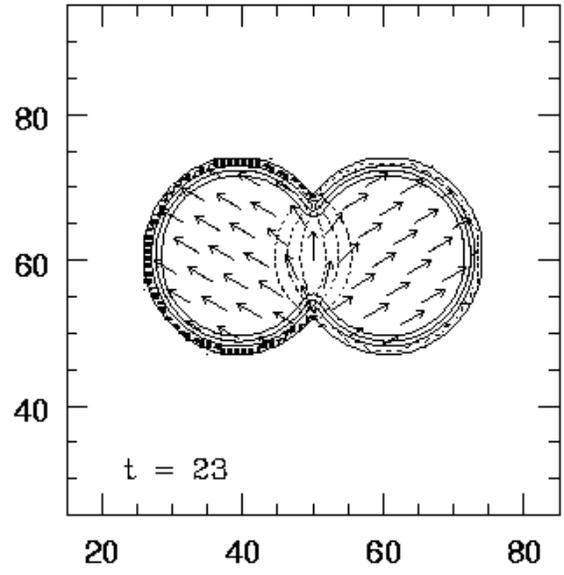
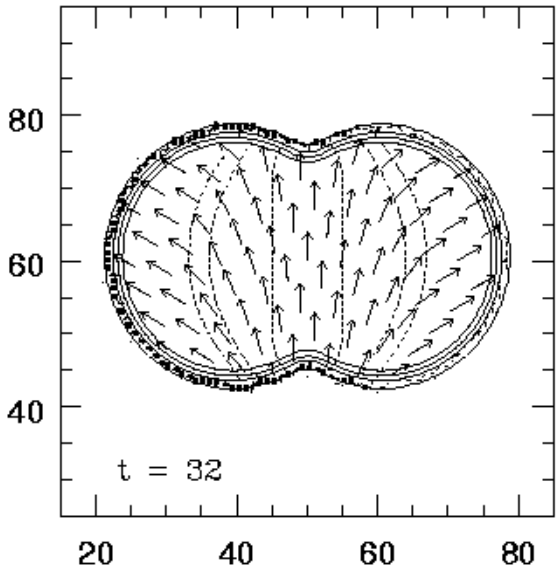
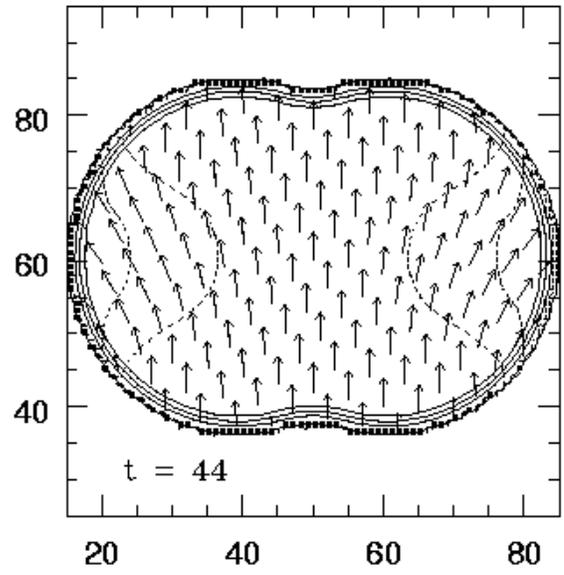
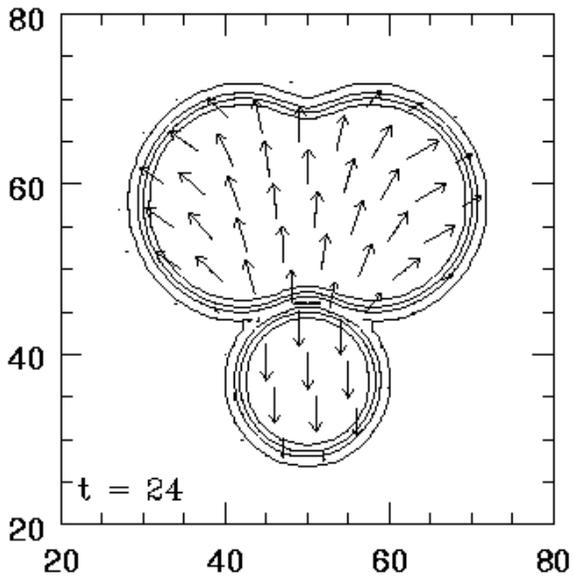
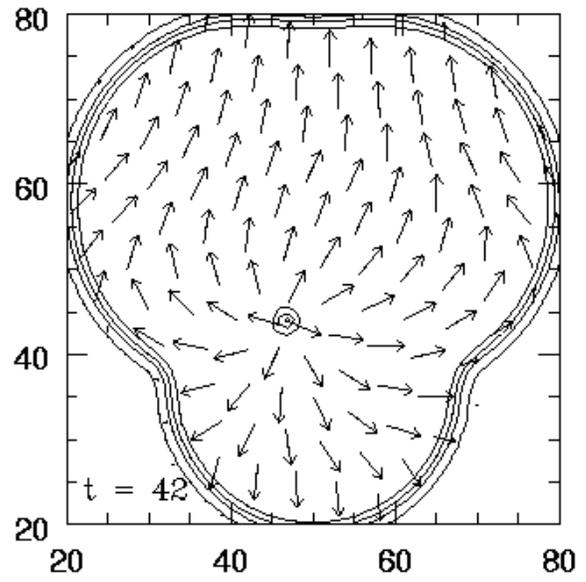
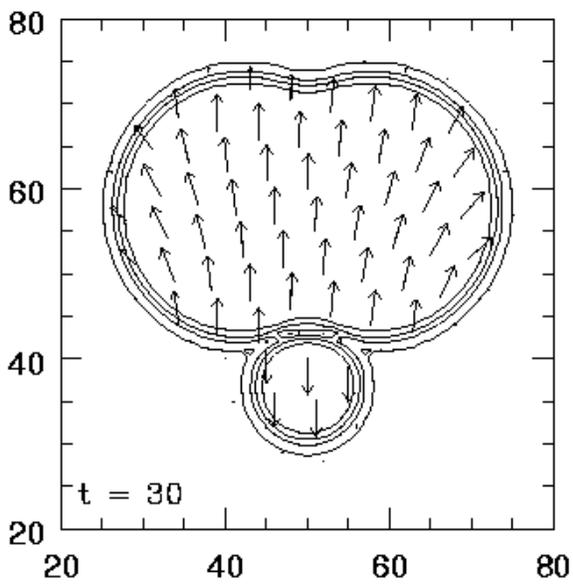
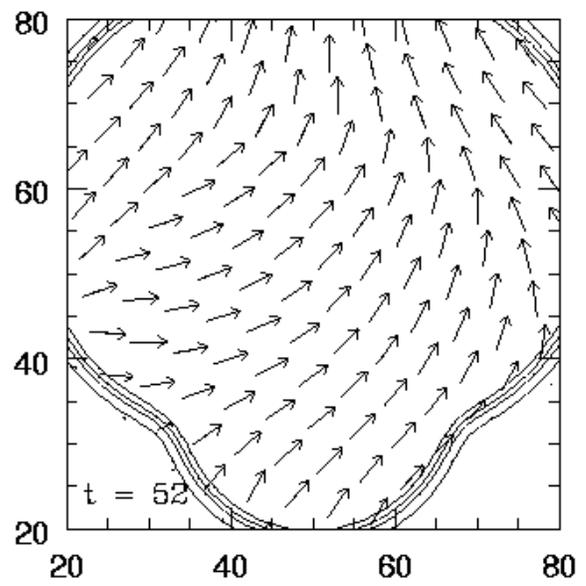
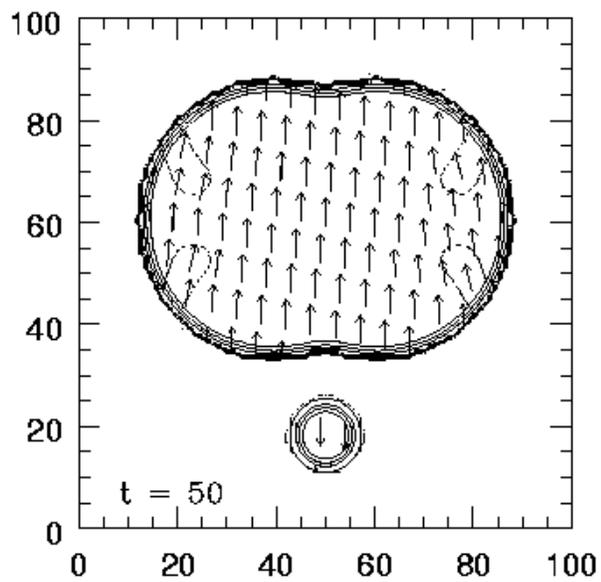
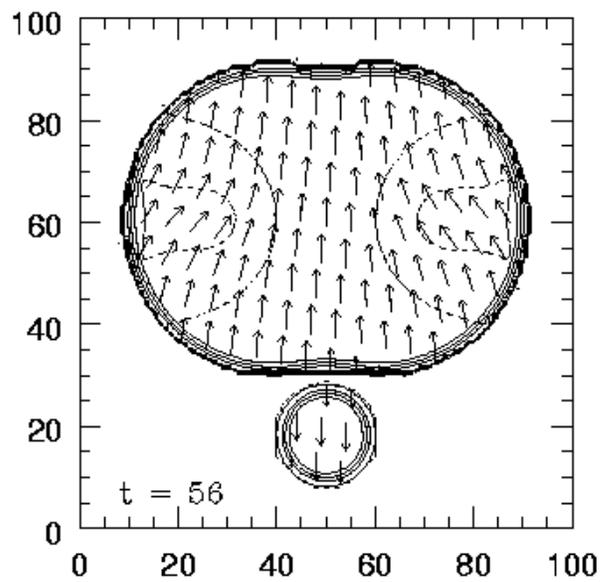
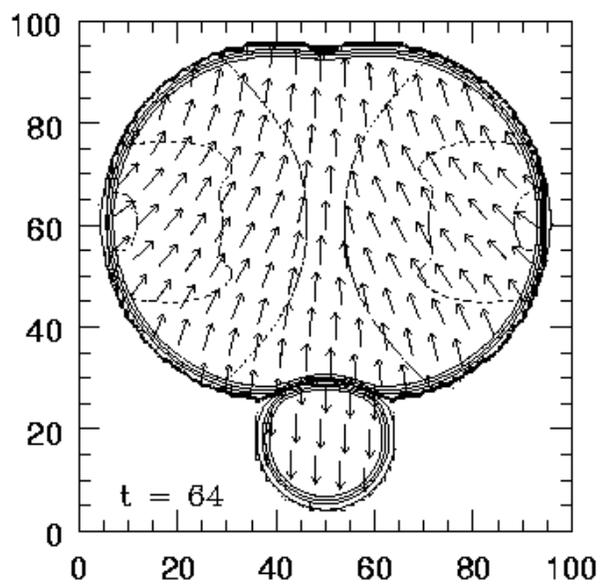
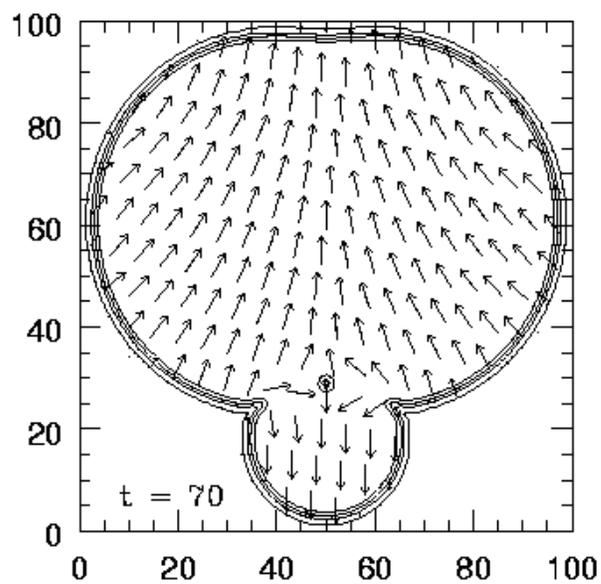
\end{document}